\documentclass[iop]{emulateapj}

\newcommand{\lya}{Ly$\alpha$}
\newcommand{\nv}{N\,{\sc v}}
\newcommand{\siii}{Si\,{\sc ii}}
\newcommand{\oi}{O\,{\sc i}}
\newcommand{\cii}{C\,{\sc ii}}
\newcommand{\siiv}{Si\,{\sc iv}}

\newcommand{\mgii}{Mg\,{\sc ii}}

\begin{document}

\title{DISCOVERY OF EIGHT $z\sim6$ QUASARS IN THE SLOAN DIGITAL SKY SURVEY 
OVERLAP REGIONS}

\author{Linhua Jiang\altaffilmark{1,2},
Ian D. McGreer\altaffilmark{3}, Xiaohui Fan\altaffilmark{3},
Fuyan Bian\altaffilmark{4,6}, Zheng Cai\altaffilmark{3}, 
Benjamin Cl{\'e}ment\altaffilmark{3}, Ran Wang\altaffilmark{1},
and Zhou Fan\altaffilmark{5}}

\altaffiltext{1}{Kavli Institute for Astronomy and Astrophysics, Peking 
   University, Beijing 100871, China; jiangKIAA@pku.edu.cn}
\altaffiltext{2}{School of Earth and Space Exploration, Arizona State
   University, Tempe, AZ 85287-1504, USA}
\altaffiltext{3}{Steward Observatory, University of Arizona,
   933 North Cherry Avenue, Tucson, AZ 85721, USA}
\altaffiltext{4}{Research School of Astronomy and Astrophysics, Australian
   National University, Weston Creek, ACT 2611, Australia}
\altaffiltext{5}{Key Laboratory of Optical Astronomy, National Astronomical
   Observatories, Chinese Academy of Sciences, Beijing 100012, China}
\altaffiltext{6}{Stromlo Fellow}

\begin{abstract}

We present the discovery of eight quasars at $z\sim6$ identified in the Sloan 
Digital Sky Survey (SDSS) overlap regions. Individual SDSS imaging runs have 
some overlap with each other, leading to repeat observations over an area 
spanning $>$4000 deg$^2$ (more than 1/4 of the total footprint). These 
overlap regions provide a unique dataset that allows us to select 
high-redshift quasars more than 0.5 mag fainter in the $z$ band than those 
found with the SDSS single-epoch data. Our quasar candidates were first 
selected as $i$-band dropout objects in the SDSS imaging database. We then 
carried out a series of follow-up observations in the optical and near-IR 
to improve photometry, remove contaminants, and identify quasars. 
The eight quasars reported here were discovered in a pilot study utilizing
the overlap regions at high galactic latitude ($|b|>30$ deg). 
These quasars span a redshift range of $5.86<z<6.06$ and a flux range of 
$19.3<z_{\rm AB}<20.6$ mag. Five of them are fainter than $z_{\rm AB}=20$ mag, 
the typical magnitude limit of $z\sim6$ quasars used for the SDSS 
single-epoch images. In addition, we recover eight previously known quasars 
at $z\sim6$ that are located in the overlap regions. These results validate
our procedure for selecting quasar candidates from the overlap regions and
confirming them with follow-up observations, and provide 
guidance to a future systematic survey over all SDSS imaging regions with
repeat observations.

\end{abstract}

\keywords
{cosmology: observations --- quasars: general --- quasars: emission lines}

\section{INTRODUCTION}

High-redshift quasars provide a powerful tool to study the early universe.
The first $z\sim6$ quasars \citep[e.g.][]{fan01,fan03,fan04} were found 
by the Sloan Digital Sky Survey \citep[SDSS;][]{york00}. Most of these quasars
are very luminous ($M_{1450}<-26$ mag). Deep near-IR spectroscopy has shown 
that these quasars harbor billion-solar-mass black holes and emit near the 
Eddington limit, suggesting the rapid growth of central black holes in the 
early epoch \citep[e.g.][]{jiang07,kurk07}. Their broad emission lines exhibit 
solar or supersolar metallicity, indicating that vigorous star formation and 
element enrichment have occurred in the host galaxies 
\citep[e.g.][]{jiang07,kurk07,derosa11}. Their optical absorption spectra show 
that the state of the intergalactic medium (IGM) at $z\sim6$ is close to the 
reionization epoch \citep[e.g.][]{fan06b,car10,mcg11}.
Furthermore, the mid/far-infrared, mm/sub-mm, and radio observations of these
quasars have provided rich information about dust emission and star formation 
in their host galaxies \citep[e.g.][]{wal09,jiang10,wang11,wang13}.
Therefore, $z\ge6$ quasars are essential to understanding black hole 
accretion, galaxy evolution, and the IGM state in the first billion years of
cosmic time.

In recent years, more than 70 quasars at $z\sim6$ have been discovered.
The SDSS pioneered quasar studies at these redshifts, followed by the Canada-France 
High-redshift Quasar Survey \citep[CFHQS;][]{wil05} and the UKIRT Infrared 
Deep Sky Survey \citep[UKIDSS;][]{war07}. To date $\sim$30 quasars have been
discovered in the main imaging survey 
\citep[e.g.][]{fan01,fan03,fan04,fan06a} 
and the Stripe 82 deep survey \citep{jiang08,jiang09}. The UKIDSS has found
several quasars \citep{ven07,mort09,mort11}, including the most distant quasar 
known at $z=7.08$ \citep{mort11}. The CFHQS produced 20 quasars down to a 
fainter luminosity limit \citep{wil07,wil09,wil10}. Most recently, two large 
surveys, the Panoramic Survey Telescope \& Rapid Response System 1 
(Pan-STARRS1) and the Visible and Infrared Survey Telescope for Astronomy
(VISTA) Kilo-Degree Infrared Galaxy (VIKING) survey have started to yield 
high-redshift quasars. Pan-STARRS1 is sensitive mainly to brighter quasars
than SDSS but over a much larger area, and $\ga$10 new quasars at $z\sim6$ 
have been published from this survey \citep{morg12,ban14}, including 
three quasars at $6.5<z<6.7$ \citep{ven15}. VIKING data have also been used
to discover three quasars at $z>6.5$ \citep{ven13}. The number of 
high-redshift known quasars is increasing steadily.

In this paper we introduce a new method for finding $z\sim6$ quasars by 
using regions with overlapping imaging in the SDSS. The SDSS imaging survey 
was conducted in drift scan mode along great circles, and the imaging runs 
usually overlap with each other due to the survey geometry and strategy. 
This results in multiple observations of individual sources within the overlap
regions. About one-quarter of the sky area covered by the SDSS has multiple 
imaging coverage. These overlap regions allow us to select quasars that are 
fainter than those found in the SDSS single-epoch images. They ensure that 
quasar candidates are free of spurious detections (such as cosmic rays), which 
is particularly useful for objects with single-band ($z$-band in this paper) 
detections. Here we report eight new quasars found in our first search of
$z\sim6$ quasars in the overlap regions. These quasars are $\sim$0.5 mag 
fainter than the quasars selected from SDSS single-epoch data. 

The layout of the paper is as follows. In Section 2 we briefly describe the
SDSS overlap regions. In Section 3 we introduce our quasar selection 
procedure and follow-up observations. We present the new quasars in Section 4,
and summarize the paper in Section 5. Throughout the paper SDSS magnitudes 
are reported on the AB system instead of the asinh system \citep{lup99}, 
and all near-IR magnitudes are on the Vega system. We use a 
$\Lambda$-dominated 
flat cosmology with $H_0=70$ km s$^{-1}$ Mpc$^{-1}$, $\Omega_{m}=0.3$, and
$\Omega_{\Lambda}=0.7$.

\section{SDSS OVERLAP REGIONS}

In this section we describe the SDSS overlap regions, the regions that were 
scanned by two or more SDSS imaging runs. The SDSS imaging survey was carried 
out in drift-scan mode using a 142 mega-pixel camera \citep{gunn98,gunn06}.
An SDSS run (strip) consists of 6 parallel scanlines (camera columns) for each 
of five $ugriz$ bands \citep{fuk96}. The scanlines are $13\farcm5$ 
wide with gaps of roughly the same width, so two interleaving strips make a 
stripe. SDSS scanlines are divided into fields, and a field is the union of 
five $ugriz$ frames covering the same region of sky. Processing of individual
runs includes assessments of the photometric quality 
\citep{ivezic04,hogg01}.

The SDSS imaging runs generally overlap each other, due to the survey
geometry and strategy \citep{stoughton02}. 
The imaging survey in drift-scan mode was along great
circles, and had two common poles at [R.A., Decl.] = 
[$95\degr$, $0\degr$] and [$270\degr$, $0\degr$]. The fields overlap more 
substantially when they approach the survey poles (Figure 1). In addition, the 
two interleaving strips that make any stripe slightly overlap, leading to
duplicate observations in a small area.
Furthermore, if the quality of a run, or part of a run, did not meet the
SDSS standard criteria, the relevant region of this run was re-observed,
yielding duplicate observations in this region. Obviously, a combination of 
the above reasons may result in more than two observations in some regions.

\begin{figure}
\plotone{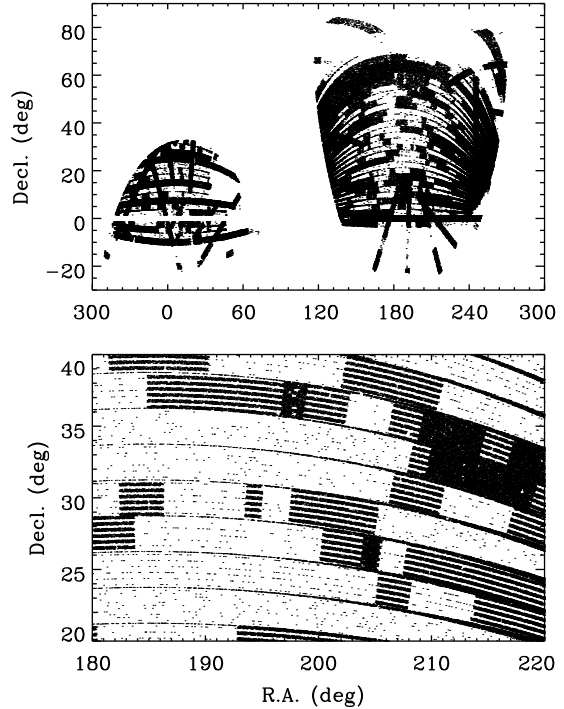}
\caption{Upper panel: coverage of the SDSS overlap regions at high galactic
latitude $|b|>30$. Stripe 82 has been excluded. The data points represent
bright ($19.0<r<19.1$) point sources with at least two SDSS observations.
Lower panel: a closer look at an area of R.A. = $180\degr$--$220\degr$ and
Decl. = $20\degr$--$40\degr$. The data points represent point sources with
$19<r<20$. We can clearly see individual scanlines.}
\end{figure}

The upper panel in Figure 1 shows the coverage of the SDSS overlap regions at 
high galactic latitude $|b|>30\degr$. The SDSS deep stripe, Stripe 82, has 
been excluded, because it has much deeper co-added data 
\citep{jiang14,annis14}. The 
data points represent bright ($19.0<r<19.1$) point sources with at least two 
observations. In the lower panel of the figure, the coverage map is zoomed in 
to an area of R.A. = $180\degr$--$220\degr$ and Decl. = $20\degr$--$40\degr$, 
where the data points are point sources with $19<r<20$.
We estimated the total area of the overlap regions in the upper panel by
calculating the area occupied by the actual data points. The sources we used
are point sources with $19<r<21$. We divided the R.A.--Decl. plane into small
grid elements, where each element measures $0.05\degr$ at a side.
If a grid element contains more than one object, it was assumed to be covered
by the overlap regions. The total area by summing the area of all selected 
grid elements is 3700 square degrees. This rough estimate is accurate to
roughly 10\%, and could be improved by using more sources and by refining 
grid elements. We leave this for a future work. The total area of all SDSS 
overlap regions, including low galactic latitude $|b|<30\degr$, is more than 
4000 square degrees.

For a given source in the overlap regions, the SDSS pipeline selects one 
detection as a `primary' detection from a `primary' run/field. Additional 
detections are classified as 
`secondary'. The primary runs/fields are generally deeper due to better image 
quality in terms of seeing, sky background, atmospheric transparency, etc. 
Figure 2 illustrates the difference between primary and
secondary detections in the $z$ band. The sources used in the figure are point 
sources with $z<21$ mag, selected from the overlap regions displayed in the 
lower panel of Figure 1 (R.A. = $180\degr$--$220\degr$ and Decl. = 
$20\degr$--$40\degr$). The upper panel of Figure 2 shows photometric errors
versus $z$-band magnitude. The shaded regions include central 68\% of the 
sources at any magnitude. 
They indicate that primary detections have slightly better
data quality than secondary detections. This is also suggested by the three
lower panels, which show the distribution of photometric errors in three
different magnitude bins. Compared to primary detections, the distribution of errors for the secondary detections has a longer tail towards greater errors. 
Therefore, a secondary detection does not always significantly improve the 
photometry of a source. However, 
it generally rules out the possibility that its primary detection is a false 
detection. This is particularly important for single-band detected sources,
such as $z\ge6$ quasars. 

\begin{figure}
\plotone{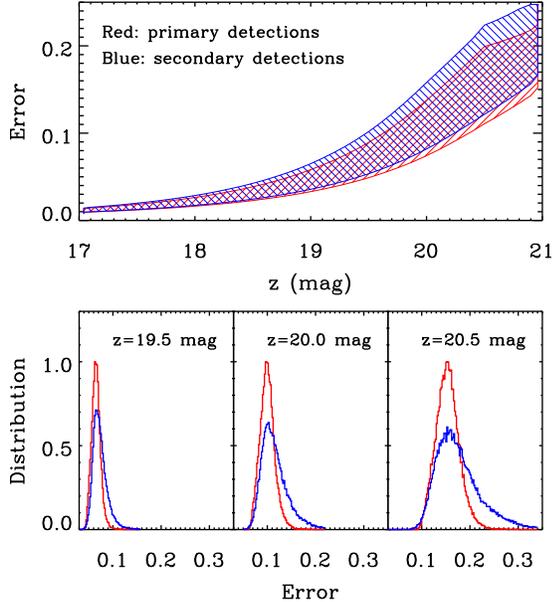}
\caption{Comparison between primary and secondary detections in the $z$ band.
The sources used in the figure are point sources with $z<21$ mag, selected
from the overlap regions displayed in the lower panel of Figure 1 (R.A. =
$180\degr$--$220\degr$ and Decl. = $20\degr$--$40\degr$).
The upper panel shows photometric errors versus $z$-band photometry. The
shaded regions include central 68\% sources at any magnitude. They indicate
that primary detections have slightly better data quality than secondary
detections. The three lower panels show the distribution of photometric errors
at three different magnitudes. Compared to primary detections, secondary
detections have longer distribution tails towards larger errors.}
\end{figure}

\section{QUASAR CANDIDATE SELECTION AND FOLLOW-UP OBSERVATIONS}

In this section we present the details of our candidate selection and 
spectroscopic identification of 
$z\sim6$ quasars in the SDSS overlap regions. The selection procedure is
similar to the procedures that were used to select SDSS $z\sim6$ quasars in
previous studies \citep[e.g.][]{fan01,fan03,jiang08,jiang09}. The procedure 
consists of several steps as shown in the following subsections. 

\subsection{Selection of $i$-band Dropout Objects}

We start with SDSS Data Release 9 (DR9)
`primary' sources at high galactic latitude $|b|>30\degr$,
by using the `photoPrimary' table with the SDSS Query/CasJobs online server,
excluding Stripe 82. The `photoPrimary' table 
includes both point and extended sources. Although distant quasars are 
point-like objects in ground-based images, faint point-like objects could be 
mis-classified as extended objects (and vice versa); thus we include 
extended objects. We then remove sources with 
critical SDSS processing flags such as BRIGHT, EDGE and SATUR. For each 
source, we searched for `neighbors' (`secondary' detections from different 
runs) within a $0\farcs5$ radius. All matched pairs are considered to be  
multiple observations of the same object. From these dual observations we select 
$i$-band dropout objects using the following color cuts,
\begin{eqnarray}
	g>24.0, \\
	r-z>3.0, \\
   i-z>2.2, \\
   {\rm and}\ \sigma_z<0.155.
\end{eqnarray}
These selection criteria are applied to the individual observations, i.e.,
each of the dual observations must meet the criteria except criterion 4. We 
require that the primary detections meet criterion 4 (at least 
$7\sigma$ detections), and the secondary detections should be detected at a 
significance of $>5\sigma$. 
In this first step, we 
obtain a total of $\sim$550 $i$-band dropout objects. 

In the previous search of luminous $z\sim6$ quasars with the SDSS single-epoch 
images (see Fan et al. papers), the magnitude limit was roughly $z=20$  
($10\sigma$ detections). Below $10\sigma$, the number of 
contaminants in $i$-dropout selection increases dramatically. From single-epoch 
images, one would get tens of thousands of $i$-dropout candidates down to 
a $7\sigma$ threshold, where the vast majority of them are spurious objects.
Using duplicate observations ensures that most of the 
$i$-dropout objects we select are physical sources.

We visually inspect the dropouts in the SDSS $ugriz$ images, and remove 
obvious contaminants, such as suspicious detections close to very bright 
stars and objects that are apparently extended. We reject about
one-third of the 550 objects in this step.
We then remove previously known objects, such as L/T dwarfs and $z\sim6$ 
quasars. We match our objects to the list of known L/T dwarfs from 
{\tt DwarfArchives.org}, and find 51 known dwarfs. We also match the 
objects to the list of $i$-dropouts that have been spectroscopically
identified as non-quasars during our previous searches for $z\sim6$ quasars.
In addition, we find that 8 objects are previously
discovered quasars at $z\sim6$. These quasars are summarized in Table 1. 
They are generally brighter than $z=20$ mag. After removing these objects,
about 250 objects remain in our candidate list.

We match the remaining objects to the 2MASS catalog \citep{skr06} and to 
the catalog of the UKIRT Infrared Deep Sky Survey \citep[UKIDSS;][]{law07}.
We obtain 10 matches from the 2MASS, and 53 matches from the UKIDSS,
providing near-IR photometry for 63 objects. 
The majority are then rejected as quasar candidates based on their location 
in the $z-J$ versus $i-z$ color-color diagram (see section 3.3). This leaves
a total of 190 $i$-dropout candidates selected for additional observations.

\begin{deluxetable*}{ccccl}
\tablecaption{Previously Known Quasars in the SDSS Overlap Regions}
\tablehead{\colhead{Quasar (SDSS)} & \colhead{Redshift} & \colhead{$i$ (mag)} &
   \colhead{$z$ (mag)} & \colhead{Reference} }
\startdata
J000239.40+255034.8 & 5.80 & 21.49$\pm$0.12 & 18.94$\pm$0.05 & \citet{fan04} \\
                    &      & 21.39$\pm$0.13 & 18.88$\pm$0.05 &  \\
J084119.52+290504.4 & 5.96 & 23.30$\pm$0.66 & 19.78$\pm$0.13 & \citet{goto06} \\
                    &      & 22.47$\pm$0.28 & 19.78$\pm$0.09 &  \\
J103027.09+052455.0 & 6.28 & 22.93$\pm$0.41 & 19.59$\pm$0.08 & \citet{fan01} \\
                    &      & 23.14$\pm$0.42 & 19.91$\pm$0.11 &  \\
J104433.04--012502.1& 5.80 & 21.58$\pm$0.20 & 18.99$\pm$0.07 & \citet{fan00} \\
                    &      & 21.62$\pm$0.16 & 19.09$\pm$0.07 &  \\
J131911.29+095051.3 & 6.13 & 21.79$\pm$0.33 & 19.35$\pm$0.14 & \citet{mort09} \\
                    &      & 22.69$\pm$0.28 & 20.07$\pm$0.10 &  \\
J141111.28+121737.3 & 5.93 & 22.91$\pm$0.35 & 19.55$\pm$0.08 & \citet{fan04} \\
                    &      & 23.49$\pm$0.42 & 19.56$\pm$0.07 &  \\
J162331.81+311200.6 & 6.22 & $>24$          & 19.64$\pm$0.10 & \citet{fan04} \\
                    &      & 24.47$\pm$0.85 & 20.06$\pm$0.15 &  \\
J163033.90+401209.7 & 6.05 & 23.42$\pm$0.39 & 20.33$\pm$0.12 & \citet{fan03} \\
                    &      & 22.69$\pm$0.27 & 20.14$\pm$0.11 &  \\
\enddata
\tablecomments{The multiple entries for the $i$ and $z$ band photometry are
from SDSS duplicate observations.}
\end{deluxetable*}

\subsection{Improving the $i$ and $z$-band Photometry}

The next step is to improve the $i$ and $z$-band photometry. The majority of 
the objects have $z$-band detections weaker than $10\sigma$. For these 
objects we obtained deeper $i$ and $z$ band images using the wide-field 
optical imager 90Prime on the 2.3m Bok 
telescope. The typical integration time was 5 min per object (single exposure) 
in the $i$ band and 3 min per object (single exposure) in the $z$ band. 
The observing conditions for many objects were poor-to-moderate with 
significant moonlight, poor seeing, or non-photometric transparency, 
because most of them were observed as backup targets for other programs. 
To date we have observed roughly 110 objects.

The 90Prime images were reduced in a standard fashion using our own {\tt IDL} 
routines. First we made master bias and flat images from bias and flat images taken 
in the same night. A bad pixel mask was created from the flat image.
Then science images were overscan and bias corrected and flat-fielded. 
Saturated pixels and bleeding trails were identified, and incorporated (along
with the bad mask) into weight images. Next the first-round sky subtraction 
was performed by fitting a low-order 2D polynomial function to the background.
The 90Prime CCDs are thin chips, so they produce strong fringing in the $i$
and $z$ bands. We removed fringes in two iterations. Then another round of sky
subtraction was performed. 
We detected objects using {\tt SExtractor} \citep{ber96} and calculated 
astrometric solutions using {\tt SCAMP} \citep{ber06} by matching objects to 
the SDSS. Photometry was measured within an aperture (diameter) size of 8 
pixels ($\sim3\farcs6$) and calibrated by matching a large number of nearby 
bright point sources to the SDSS.

With the improved $i$ and $z$ band photometry, we rejected objects with 
relatively blue $i-z$ colors. The 90Prime $i$ and $z$ filters are slightly 
redder than the SDSS $i$ and $z$ filters. Depending the strength of \lya\ 
emission, quasars at $5.7<z<6.0$ could have much bluer $i-z$ colors in the 
90Prime system. We rejected objects with $i-z<1.8$ (in the 90Prime system), 
instead of $i-z<2.2$ (in the SDSS filter system) that was used in our previous 
studies. This is similar to the color cut used by 
the CFHQS \citep[e.g.,][]{wil07}. 

\subsection{Near-IR Photometry and Optical Spectroscopy}

The next step is to obtain near-IR ($J$ and/or $Y$ band) photometry. We 
excluded objects that already have $J$-band photometry from the 2MASS 
or the UKIDSS (section 3.1). Near-IR observations for the other objects were 
obtained with the MMT SWIRC \citep{bro08} and the KPNO 4m NEWFIRM. 
The objects were observed as backup 
targets for other programs, so the observing conditions were usually poor.
The exposure time per object was between 3 and 10 min. We have observed
roughly 40 objects to date. The near-IR images were 
reduced using standard IRAF\footnote{IRAF
is distributed by the National Optical Astronomy Observatory, which is
operated by the Association of Universities for Research in Astronomy (AURA)
under cooperative agreement with the National Science Foundation.} routines.
The near-IR photometry is used to separate quasars from late-type dwarf 
stars, which are the primary contaminants in $i$-dropout searches, as 
mentioned earlier. Stars can be rejected based on the $z-J$ versus $i-z$ 
color-color diagram; a more detailed description can be found in 
\citet{jiang08,jiang09}. 
In short, we use the following criterion to further select quasar candidates,
\begin{equation}
z-J<0.5(i-z)+0.5.
\end{equation}

In the final step we conducted optical spectroscopy to identify quasar
candidates using the MMT Red Channel spectrograph. We have observed
about 10 candidates, preferentially choosing candidates with UKIDSS
photometry in the $Y$ and $J$ bands. The $Y-J$ color is one of the best criteria for separating $z\sim6$ quasars L and T dwarfs.
The MMT observations were made
in long-slit mode with a spectral resolution of $\sim$10 \AA. The exposure
times for each target were 5--10 min, sufficient to classify our
candidates as quasars under normal weather conditions.
If a target was identified as a quasar immediately after its first spectrum 
was obtained, additional 5 or 10 min exposures were taken to improve 
the spectral quality. Two quasars were identified with poor observing 
conditions, and their additional spectra were taken one year later.
The spectra of one quasar (SDSS J0842+1218) was obtained with Keck/ESI.
The spectra were reduced using standard IRAF routines. In total
we identified eight new quasars from the optical spectroscopic observations.

\begin{deluxetable*}{ccccccc}
\tablecaption{Eight New Quasars in the SDSS Overlap Regions}
\tablehead{\colhead{Quasar (SDSS)} & \colhead{Redshift} & \colhead{$i$ (mag)\tablenotemark{a}} &
   \colhead{$z$ (mag)\tablenotemark{a}} & \colhead{MJD\tablenotemark{b}} & \colhead{$Y$ (mag)} & \colhead{$J$ (mag)} }
\startdata
J000825.77--062604.6 & 5.929$\pm$0.003 & 23.91$\pm$0.65 & 20.04$\pm$0.15 & 53997  &                & 19.43$\pm$0.13 \\
                     &                 & 24.04$\pm$1.03 & 19.55$\pm$0.13 & 55120  &                &                \\
                     &                 & 22.85$\pm$0.25 & 20.35$\pm$0.09 & 56561  &                &                \\
J002806.57+045725.3  & 6.04$\pm$0.03   & 22.73$\pm$0.27 & 20.30$\pm$0.14 & 54742  & 19.59$\pm$0.09\tablenotemark{c} & 19.16$\pm$0.12\tablenotemark{c} \\
                     &                 & $>$24          & 20.42$\pm$0.22 & 54764  &                &                \\
                     &                 & 24.00$\pm$0.50 & 20.49$\pm$0.10 & 56560  &                &                \\
J014837.64+060020.0  & 5.923$\pm$0.003 & 22.25$\pm$0.18 & 19.31$\pm$0.06 & 53655  & 18.91$\pm$0.06\tablenotemark{c} & 18.37$\pm$0.07\tablenotemark{c} \\
                     &                 & 22.09$\pm$0.22 & 19.35$\pm$0.08 & 53654  &                &                \\
J084229.43+121850.5  & 6.055$\pm$0.003 & 23.31$\pm$0.39 & 19.56$\pm$0.08 & 53441  &                & 18.84$\pm$0.03 \\
                     &                 & 23.43$\pm$0.72 & 19.46$\pm$0.12 & 53684  &                &                \\
J085048.25+324647.9  & 5.867$\pm$0.007 & 24.25$\pm$0.78 & 20.34$\pm$0.16 & 52664  &                & 18.90$\pm$0.10 \\
                     &                 & 23.66$\pm$0.52 & 20.39$\pm$0.15 & 52664  &                &                \\
                     &                 &                & 19.95$\pm$0.08 & 56369  &                &                \\
J120737.43+063010.1  & 6.040$\pm$0.003 & 22.70$\pm$0.29 & 20.16$\pm$0.12 & 52338  & 19.51$\pm$0.09\tablenotemark{c} & 19.35$\pm$0.14\tablenotemark{c} \\
                     &                 & 23.49$\pm$0.61 & 20.17$\pm$0.16 & 52435  &                &                \\
                     &                 & $>$23.5        & 20.39$\pm$0.11 & 56370  &                &                \\
J125757.47+634937.2  & 6.02$\pm$0.03   & 24.91$\pm$0.84 & 20.20$\pm$0.15 & 51987  & 20.39$\pm$0.16 & 19.78$\pm$0.08 \\
                     &                 & 23.55$\pm$0.54 & 20.29$\pm$0.14 & 51633  &                &                \\
                     &                 & 23.50$\pm$0.40 & 20.60$\pm$0.12 & 56394  &                &                \\
J140319.13+090250.9  & 5.86$\pm$0.03   & 22.53$\pm$0.23 & 20.17$\pm$0.14 & 52757  & 19.70$\pm$0.12\tablenotemark{c} & 19.17$\pm$0.10\tablenotemark{c} \\
                     &                 & 23.01$\pm$0.32 & 20.26$\pm$0.13 & 52345  &                &                \\
                     &                 & 22.73$\pm$0.22 & 20.48$\pm$0.11 & 56394  &                &                \\
\enddata
\tablenotetext{a}{The first two entries for each quasar were from the SDSS, and
the third entry (if there is one) was from the Bok 90Prime.}
\tablenotetext{b}{MJD for the $i$ and $z$-band photometry.}
\tablenotetext{c}{Photometry from the UKIDSS.}
\end{deluxetable*}

\section{RESULTS AND DISCUSSION}

\subsection{Discovery of Eight New Quasars at $z\sim6$}

In this section we present the discovery of eight new quasars at $z\sim6$.
Table 2 lists the coordinates, redshifts, and optical and near-IR photometry
of the new quasars. The naming convention for SDSS sources is 
SDSS JHHMMSS.SS$\pm$DDMMSS.S, and the positions are expressed in J2000.0 
coordinates. We use SDSS JHHMM$\pm$DDMM (or JHHMM in figures) for brevity.
The new quasars span a relatively narrow redshift range of $5.86<z<6.06$. The 
redshifts were mostly measured from several emission lines including \lya, 
\oi\ $\lambda1304$ (hereafter \oi), \cii\ $\lambda1335$ (hereafter \cii),
and \siiv\ $\lambda1396$ (hereafter \siiv), by fitting a Gaussian profile to 
the top $20\sim50$\% of these lines. Three quasars do not show apparent
line emission in their optical spectra, so their redshifts were measured from
the wavelength where the sharp flux decline occurs. They are discussed in more detail  
in the next subsection. The redshift errors quoted in Table 2 include the 
uncertainties from our fitting process and wavelength calibration. They do not 
include any possible velocity shifts between these lines and systemic 
redshifts. For the quasars without apparent line emission, we adopted a redshift 
error of 0.03, corresponding to the scatter in the relation between \lya\ 
redshifts and systemic redshifts \citep[e.g.][]{shen07}.
Two of the quasars (SDSS J0148+0600 and SDSS J1207+0630) were independently
discovered by S. Warren et al. (in preparation).

In Table 2, there are multiple entries for the $i$ and $z$ band photometry.
The first two entries for each quasar are the duplicate observations from the 
SDSS, and the third 
one (if present) is the improved photometry from Bok 90Prime. Figure 3 shows 
two examples of the multi-epoch observations of the 
quasars SDSS J0008--0626 and SDSS J0028+0457. In the SDSS observations 
(Epochs 1 and 2), the two quasars were not (or barely) detected in the $i$ 
band, and were very weak ($5\sigma-7\sigma$) in the $z$ band. In the Bok 
90Prime observations (Epoch 3) they are clearly detected in the $i$ band, and have 
strong detections ($\ge10\sigma$) in the $z$ band. 
Note that quasars are variable objects with a typical variability amplitude of a few
tenths of a magnitude \citep[e.g.][]{mac12}, so the difference of the 
photometry in different entries for the same quasar may partially reflect 
this variability.
SDSS J0148+0600 and SDSS J0842+1218 do not have the third entry of the $i$ and 
$z$ band photometry. They are more than $10\sigma$ detections in the SDSS 
observations, so we did not take more images for them. The other six quasars 
are weaker than $10\sigma$ detections, which was the limit in the 
previous survey of $z\sim6$ quasars with the SDSS single-epoch images.

\begin{figure}
\plotone{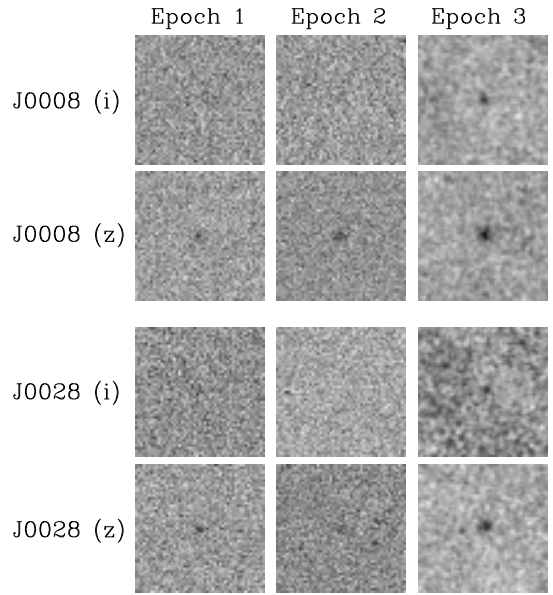}
\caption{Multi-epoch observations of SDSS J0008--0626 and SDSS J0028+0457 in
the $i$ and $z$ bands. In the SDSS observations (Epoch 1 and 2), the two
quasars were not (or barely) detected in the $i$ band, and were very weak
($5\sigma-7\sigma$) in the $z$ band. The new Bok 90Prime observations (Epoch
3) clearly detected them in the $i$ band, and have strong detections
($\ge10\sigma$) in the $z$ band.}
\end{figure}

The last two columns of Table 2 show the near-IR ($Y$ and $J$) photometry.
The $Y$ and $J$ band photometry for four quasars was obtained from the 
UKIDSS. As we mentioned in section 3.3, in this first study we preferentially 
observed candidates with UKIDSS $Y$ and $J$-band photometry.
The near-IR photometry of the other four quasars was obtained with the MMT 
SWIRC. 

Figure 4 shows the optical spectra of the quasars. As addressed in section 
3.3, one quasar (SDSS J0842+1218) was observed with Keck/ESI, and its total
integration time was 20 min. The other seven quasars were observed with the 
MMT Red Channel Spectrograph. The total integration time per object was from 
10 min to 40 min, depending on the quasar brightness and observing conditions.
Each spectrum in Figure 4 has been scaled to match the best corresponding 
$z$-band magnitude in Table 1, thereby placed on an absolute flux scale. 
These quasars show a diversity of \lya\ emission. For example, 
SDSS J0008--0626 and SDSS J0842+1218 have strong \lya\ emission 
lines. SDSS J0148+0600 and SDSS J1207+0630 are likely broad absorption line 
(BAL) quasars. 
SDSS J1257+6349 and SDSS J1403+0902 do not show any prominent \lya\ emission.
Quasars with weak emission lines like these quasars are not rare at high 
redshift. We have found a significant fraction of lineless quasars at $z\sim6$
\citep[e.g.][]{jiang09,ban14}. 

\begin{figure}
\plotone{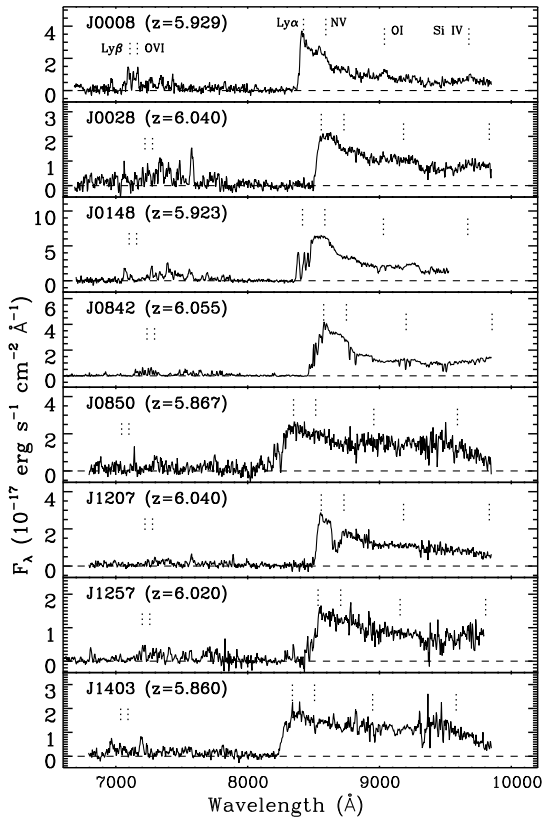}
\caption{Optical spectra of the eight quasars. The spectrum of SDSS J0842+1218
was taken from Keck/ESI. The spectra of the other quasars were taken from the
MMT Red Channel Spectrograph. Each spectrum has been scaled
to match the corresponding $z$-band magnitude in Table 1, thereby placed on an
absolute flux scale.}
\end{figure}

Table 3 shows the properties of the quasar continua and \lya\ emission lines. 
Columns 3 and 4 list $m_{1450}$ and $M_{1450}$, the apparent and absolute AB 
magnitudes of the continuum at rest-frame 1450 \AA. They were calculated by
fitting a power-law continuum $f_{\nu}\sim\nu^{\alpha_{\nu}}$ to the spectrum
of each quasar. Our optical spectra only cover a short wavelength range in the 
rest frame ($100\sim150$ \AA), and this range contains several UV emission 
lines, so we were not able to reliably measure the slopes of the continua. 
We assumed that the slope is the average quasar UV continuum slope 
$\alpha_{\nu}=-0.5$ \citep[e.g.][]{van01}. The power-law continuum was then 
normalized to match visually identified continuum windows with little 
contribution from line emission. 
After the power-law continuum was subtracted from the spectrum, we measured
the rest-frame equivalent width (EW) of the \lya\ emission line, including
the \nv\ $\lambda1240$ (hereafter \nv) line that is usually blended with \lya.
This was done by fitting a half Gaussian profile (\lya) and a full Gaussian
profile (\nv)  simultaneously to the spectrum redward of the \lya\ line center.
We ignored the weak \siii\ $\lambda1262$ emission line on the red side of \nv. 
For the quasars without prominent \lya\ emission, we simply integrated flux
(instead of Gaussian fitting) above the continuum. 
We did not correct for the flux absorbed by the \lya\ forest. The results 
are shown in Column 5 in Table 3. These measurements are crude due to 
uncertainties from continuum measurement, redshift determination, and \lya\ 
forest absorption. 

\begin{deluxetable*}{cccccc}
\tablecaption{Properties of Continua and Emission Lines}
\tablewidth{0pt}
\tablehead{\colhead{Quasar (SDSS)} & \colhead{Redshift} &
  \colhead{$m_{1450}$ (mag)} & \colhead{$M_{1450}$ (mag)} &
  \colhead{$\rm EW_{Ly\alpha+NV}$ (\AA)} }
\startdata
J0008--0626 & 5.929  &  20.63$\pm$0.09  &  --26.04$\pm$0.09  &  78   \\
J0028+0457  & 6.040  &  20.34$\pm$0.10  &  --26.38$\pm$0.10  &  17   \\
J0148+0600  & 5.923  &  19.60$\pm$0.06  &  --27.08$\pm$0.06  & $>87$ \\
J0842+1218  & 6.055  &  19.85$\pm$0.09  &  --26.85$\pm$0.09  &  44   \\
J0850+3246  & 5.867  &  19.92$\pm$0.08  &  --26.74$\pm$0.08  &  10   \\
J1207+0630  & 6.040  &  20.10$\pm$0.11  &  --26.60$\pm$0.11  &  31   \\
J1257+6349  & 6.02   &  20.56$\pm$0.12  &  --26.14$\pm$0.12  &  18   \\
J1403+0902  & 5.86   &  20.39$\pm$0.11  &  --26.27$\pm$0.11  &  8   \\
\enddata
\end{deluxetable*}

\subsection{Notes on Individual Objects}

{\it SDSS J0008--0626 (z=5.929).}
SDSS J0008--0626 has a strong \lya\ emission line. It also shows prominent 
\nv, \oi, and \siiv\ emission lines. Its redshift was measured from the \oi\ 
emission line.

{\it SDSS J0028+0457 (z=6.04).}
SDSS J0028+0457 was independently discovered by the Pan-STARRS1 survey
\citep{ban14}. It does not have prominent emission lines in the spectrum 
shown in Figure 4. The redshift measured from the weak \lya\ line peak is 
6.09, and the redshift measured from the wavelength where the sharp flux drop 
occurs is 6.04. Here we adopt 6.04. This is slightly higher than the redshift 
reported in \citet{ban14} ($z=$5.99). An interesting feature
is a strong narrow emission line at $\lambda=7580$ \AA. It exists in all three
individual MMT spectra, as well as the \citet{ban14} spectrum, so 
it is a real line feature. It could be a luminous \lya\ emitter at 
$z=5.23$, or an ionized IGM bubble at $z=5.23$. Higher resolution imaging
and spectroscopy are needed to explore the nature of this line.

{\it SDSS J0148+0600 (z=5.923).}
SDSS J0148+0600 is the brightest quasar in this sample. It was well detected
in the SDSS single-epoch images and the UKIDSS near-IR images. It is likely a 
low-ionization BAL (LoBAL) quasar, and its \lya\ emission line has been mostly 
absorbed. The EW listed in Table 3 is a lower limit. It does not have obvious 
emission lines other than \cii\ redward of \nv\ in its optical spectrum, so we 
used \cii\ to determine the redshift. This quasar was independently 
discovered by S. Warren et al. (in preparation). \citet{bec15} found that it
showed a very long Gunn-Peterson trough in a deep VLT/X-Shooter spectrum.

{\it SDSS J0842+1218 (z=6.055).}
SDSS J0842+1218 was discovered in 2007. Its optical spectrum was obtained
with Keck/ESI (20 min exposure in total). It has been observed in the near-IR 
and mid-IR wavelength ranges, and its physical properties such as metallicity
and black hole mass have been well measured \citep[e.g.][]{jiang10,derosa11}.
This quasar was initially selected from the SDSS single-epoch images, but 
also satisfies our overlap region selection criteria.
We include this quasar because it is in the overlap regions and its 
optical spectrum was never published. The redshift of this quasar measured
from the \lya\ line is 6.055. This is consistent with the redshift 6.08 
measured from its \mgii\ emission line \citep{derosa11}, given the large
scatter ($\sim0.03$) between \lya\ redshifts and systemic redshifts.

{\it SDSS J0850+3246 (z=5.867).}
SDSS J0850+3246 shows a very weak \lya\ emission line with an observed 
rest-frame EW about 10\AA. It does not show any other line emission in its
optical spectrum in Figure 4.

{\it SDSS J1207+0630 (z=6.040).}
SDSS J1207+0630 is a high-ionization BAL (HiBAL) quasar, as seen from the
strong \nv\ absorption line in the spectrum. It was also covered and detected
in the near-IR by the UKIDSS.

{\it SDSS J1257+6349 (z=6.02).}
SDSS J1257+6349 does not show any significant line emission in its optical 
spectrum in Figure 4. Its redshift was measured from the sharp flux drop at
the \lya\ line.

{\it SDSS J1403+0902 (z=5.860).}
SDSS J1403+0902 is another weak line quasar, with an observed rest-frame \lya\
EW of only 8\AA. Its redshift was measured from the sharp flux drop at
the \lya\ line. It was also covered and detected in the near-IR by the UKIDSS.

\subsection{Discussion}

The currently known SDSS quasars at $z\sim6$ were selected either from 
single-epoch imaging data encompassing more than 8000 deg$^2$ with a limiting 
depth of $z_{\rm AB}\sim20$ mag, or from the Stripe 82 multi-epoch co-added 
images over 300 deg$^2$ that are two magnitudes deeper. The SDSS overlap 
regions provide another unique dataset that is in between the SDSS 
single-epoch data and the Stripe 82 data in terms of areal coverage and image d
epth. These regions cover roughly 3700 deg$^2$ area at $|b|>30\degr$, which is 
nearly one third of the total SDSS sky 
coverage, and is an order of magnitude greater than the area of Stripe 82. 
We have demonstrated that the overlap regions can be used to efficiently 
select quasar candidates that are $\ga$0.5 mag fainter than those found in 
single-epoch images. Duplicate observations allow us to select fainter 
$z\sim6$ candidates primarily because the initial candidate selection is 
largely free of 
spurious detections. The survey limit for $z\sim6$ quasars in single-epoch 
images was roughly $z_{\rm AB}=20$ mag (or $10\sigma$ detections). 
Beyond this limit, the number of spurious detections increases dramatically, 
making follow-up observations infeasible. The multi-epoch observations in the 
overlap regions ensure that the majority of single-band ($z$-band) detections
are real.

In our first quasar search we have found eight quasars in the overlap regions.
We also recovered several known quasars that are located in the overlap 
regions. As we mentioned earlier, the follow-up observations of our 
candidates have not been systematic. They were usually observed as backup
targets in other programs, when weather conditions were poor. 
Hence these quasars do not form a statistically complete sample. 
However, they provide guidance for a future systematic quasar survey 
over the whole overlap regions. In addition, we will use the WISE mid-IR 
imaging data to improve our selection efficiency \citep[e.g.][]{wu12,car15}.

\section{SUMMARY}

In this paper we have presented the discovery of eight new quasars at 
$z\sim6$. They were found in our first high-redshift quasar search in the SDSS 
overlap regions. These regions were scanned twice or more by the SDSS imaging 
survey, allowing us to select quasars fainter than those found in 
SDSS single-epoch images. We have introduced the details of our quasar 
selection procedure, including the selection of $i$-band dropouts, the 
follow-up optical and near-IR imaging observations, and the final 
spectroscopic identifications. The eight quasars span a 
redshift range of $5.86<z<6.06$. Five of them are fainter than $z_{\rm AB}=20$ 
mag, and the faintest one is $z_{\rm AB}=20.6$ mag, $\sim$0.5 mag fainter than 
those found from SDSS single-epoch images. 
These quasars show a diversity of emission line properties in their optical 
spectra, from strong emission lines to lineless features. 
In addition, we recovered eight previously known $z\sim6$ quasars that
are also located in these overlap regions. These results indicate that the SDSS overlap 
regions can be used to efficiently select high-redshift quasars. 
We plan to carry out a systematic survey of $z\sim6$ quasars in the whole SDSS 
overlap regions at high galactic latitude.

\acknowledgments

We acknowledge the support from a 985 project at Peking University.
L.J. and Z.F. acknowledge supports from the National Natural Science 
Foundation of China (NSFC) under grants 11003021 and 11373003.
X.F. and I.D.M. acknowledge support from U.S. NSF grant AST 11-07682.
Observations reported here were obtained in part at the MMT Observatory, 
a joint facility of the University of Arizona and the Smithsonian Institution.
This work is based in part on observations at KPNO, NOAO (Prop. ID: 2014B-0099; 
PI: Linhua Jiang), which is operated by the AURA under cooperative agreement 
with the NSF. This work is based in part on data obtained as part of the UKIRT 
Infrared Deep Sky Survey. This research has benefited from the M, L, T, and Y 
dwarf compendium housed at DwarfArchives.org.

Funding for the SDSS and SDSS-II has been provided by the Alfred P. Sloan
Foundation, the Participating Institutions, the National Science Foundation,
the U.S. Department of Energy, the National Aeronautics and Space
Administration, the Japanese Monbukagakusho, the Max Planck Society, and the
Higher Education Funding Council for England. The SDSS Web Site is
http://www.sdss.org/.
The SDSS is managed by the Astrophysical Research Consortium for the
Participating Institutions. The participating institutions are the American
Museum of Natural History, Astrophysical Institute Potsdam, University of
Basel, University of Cambridge, Case Western Reserve University, University
of Chicago, Drexel University, Fermilab, the Institute for Advanced Study, the
Japan Participation Group, Johns Hopkins University, the Joint Institute for
Nuclear Astrophysics, the Kavli Institute for Particle Astrophysics and
Cosmology, the Korean Scientist Group, the Chinese Academy of Sciences
(LAMOST), Los Alamos National Laboratory, the Max-Planck-Institute for
Astronomy (MPIA), the Max-Planck-Institute for Astrophysics (MPA), New Mexico
State University, Ohio State University, University of Pittsburgh, University
of Portsmouth, Princeton University, the United States Naval Observatory, and
the University of Washington.

{\it Facilities:}
\facility{Bok (90Prime)},
\facility{MMT (SWIRC, Red Channel spectrograph)},
\facility{Mayall (NEWFIRM)},
\facility{Keck (ESI)}


\begin{thebibliography}{}
\bibitem[Annis et al.(2014)]{annis14} Annis, J., Soares-Santos, 
	M., Strauss, M.~A., et al.\ 2014, \apj, 794, 120 
\bibitem[Ba{\~n}ados et al.(2014)]{ban14} Ba{\~n}ados, E., 
	Venemans, B.~P., Morganson, E., et al.\ 2014, \aj, 148, 14 
\bibitem[Becker et al.(2001)]{bec01} Becker, R.~H., Fan, X., 
	White, R.~L., et al.\ 2001, \aj, 122, 2850 
\bibitem[Becker et al.(2015)]{bec15} Becker, G.~D., Bolton, 
	J.~S., Madau, P., et al.\ 2015, \mnras, 447, 3402
\bibitem[Bertin \& Arnouts(1996)]{ber96} Bertin, E., \& Arnouts, S.\ 1996,
   \aaps, 117, 393
\bibitem[Bertin(2006)]{ber06} Bertin, E.\ 2006, in ASP Conf. Ser. 351, 
	Astronomical Data Analysis Software and Systems XV, ed. C. Gabriel, 
	C. Arviset, D. Ponz, \& E. Solano (San Francisco, CA: ASP), 112 
\bibitem[Brown et al.(2008)]{bro08} Brown, W.~R., McLeod, B.~A., Geary, 
	J.~C., \& Bowsher, E.~C.\ 2008, \procspie, 7014, 90
\bibitem[Carilli et al.(2010)]{car10} Carilli, C.~L., Wang, R., Fan, X., 
	et al.\ 2010, \apj, 714, 834
\bibitem[Carnall et al.(2015)]{car15} Carnall, A.~C., Shanks, T., 
	Chehade, B., et al.\ 2015, arXiv:1502.07748 
\bibitem[De Rosa et al.(2011)]{derosa11} De Rosa, G., Decarli, R., 
	Walter, F., et al.\ 2011, \apj, 739, 56 
\bibitem[Fan et al.(2000)]{fan00} Fan, X., White, R.~L., 
	Davis, M., et al.\ 2000, \aj, 120, 1167 
\bibitem[Fan et al.(2001)]{fan01} Fan, X., Narayanan, V.~K., 
	Lupton, R.~H., et al.\ 2001, \aj, 122, 2833 
\bibitem[Fan et al.(2003)]{fan03} Fan, X., Strauss, M.~A., 
	Schneider, D.~P., et al.\ 2003, \aj, 125, 1649
\bibitem[Fan et al.(2004)]{fan04} Fan, X., Hennawi, J.~F., 
	Richards, G.~T., et al.\ 2004, \aj, 128, 515
\bibitem[Fan et al.(2006a)]{fan06a} Fan, X., Strauss, M.~A., 
	Richards, G.~T., et al.\ 2006, \aj, 131, 1203 
\bibitem[Fan et al.(2006b)]{fan06b} Fan, X., Strauss, M.~A., 
	Becker, R.~H., et al.\ 2006, \aj, 132, 117 
\bibitem[Fukugita et al.(1996)]{fuk96} Fukugita, M., Ichikawa, T., Gunn,
   J.~E., et al.\ 1996, \aj, 111, 1748
\bibitem[Goto(2006)]{goto06} Goto, T.\ 2006, \mnras, 371, 769
\bibitem[Gunn et al.(1998)]{gunn98} Gunn, J.~E., Carr, M., Rockosi, C.,
   et al.\ 1998, \aj, 116, 3040
\bibitem[Gunn et al.(2006)]{gunn06} Gunn, J.~E., Siegmund, W.~A., Mannery,
   E.~J., et al.\ 2006, \aj, 131, 2332
\bibitem[Hogg et al.(2001)]{hogg01} Hogg, D.~W., Finkbeiner, 
D.~P., Schlegel, D.~J., \& Gunn, J.~E.\ 2001, \aj, 122, 2129 
\bibitem[Ivezi{\'c} et al.(2004)]{ivezic04} Ivezi{\'c}, {\v Z}., 
Lupton, R.~H., Schlegel, D., et al.\ 2004, Astronomische Nachrichten, 325, 
583 
\bibitem[Jiang et al.(2007)]{jiang07} Jiang, L., Fan, X., Vestergaard, M., 
	et al.\ 2007, \aj, 134, 1150 
\bibitem[Jiang et al.(2008)]{jiang08} Jiang, L., Fan, X., Annis, J.,
   et al.\ 2008, \aj, 135, 1057
\bibitem[Jiang et al.(2009)]{jiang09} Jiang, L., Fan, X., Bian, F.,
   et al.\ 2009, \aj, 138, 305
\bibitem[Jiang et al.(2010)]{jiang10} Jiang, L., Fan, X., Brandt, W.~N., 
	et al.\ 2010, \nat, 464, 380 
\bibitem[Jiang et al.(2014)]{jiang14} Jiang, L., Fan, X., Bian, F., 
	et al.\ 2014, \apjs, 213, 12 
\bibitem[Kurk et al.(2007)]{kurk07} Kurk, J.~D., Walter, F., Fan, X., 
	et al.\ 2007, \apj, 669, 32 
\bibitem[Lawrence et al.(2007)]{law07} Lawrence, A., Warren, S.~J.,
   Almaini, O., et al.\ 2007, \mnras, 379, 1599
\bibitem[Lupton et al.(1999)]{lup99} Lupton, R.~H., Gunn, 
J.~E., \& Szalay, A.~S.\ 1999, \aj, 118, 1406 
\bibitem[MacLeod et al.(2012)]{mac12} MacLeod, C.~L., 
	Ivezi{\'c}, {\v Z}., Sesar, B., et al.\ 2012, \apj, 753, 106
\bibitem[McGreer et al.(2011)]{mcg11} McGreer, I.~D., Mesinger, A., \& 
	Fan, X.\ 2011, \mnras, 415, 3237
\bibitem[Morganson et al.(2012)]{morg12} Morganson, E., De Rosa, G.,
   Decarli, R., et al.\ 2012, \aj, 143, 142
\bibitem[Mortlock et al.(2009)]{mort09} Mortlock, D.~J., Patel, M., 
	Warren, S.~J., et al.\ 2009, \aap, 505, 97
\bibitem[Mortlock et al.(2011)]{mort11} Mortlock, D.~J., Warren, S.~J., 
	Venemans, B.~P., et al.\ 2011, \nat, 474, 616
\bibitem[Shen et al.(2007)]{shen07} Shen, Y., Strauss, M.~A., Oguri, M., 
	et al.\ 2007, \aj, 133, 2222
\bibitem[Skrutskie et al.(2006)]{skr06} Skrutskie, M.~F., Cutri, R.~M.,
   Stiening, R., et al.\ 2006, \aj, 131, 1163
\bibitem[Stoughton et al.(2002)]{stoughton02} Stoughton, C., 
Lupton, R.~H., Bernardi, M., et al.\ 2002, \aj, 123, 485 
\bibitem[Vanden Berk et al.(2001)]{van01} Vanden Berk, D. E., et al. 2001,
  \aj, 122, 549
\bibitem[Venemans et al.(2007)]{ven07} Venemans, B.~P., McMahon, R.~G., 
	Warren, S.~J., et al.\ 2007, \mnras, 376, L76
\bibitem[Venemans et al.(2013)]{ven13} Venemans, B.~P., Findlay, J.~R., 
	Sutherland, W.~J., et al.\ 2013, \apj, 779, 24
\bibitem[Venemans et al.(2015)]{ven15} Venemans, B.~P., Ba{\~n}ados, E., 
	Decarli, R., et al.\ 2015, \apj, 801, 11
\bibitem[Walter et al.(2009)]{wal09} Walter, F., Riechers, 
	D., Cox, P., et al.\ 2009, \nat, 457, 699
\bibitem[Wang et al.(2011)]{wang11} Wang, R., Wagg, J., Carilli, C.~L., 
	et al.\ 2011, \aj, 142, 101 
\bibitem[Wang et al.(2013)]{wang13} Wang, R., Wagg, J., Carilli, C.~L., 
	et al.\ 2013, \apj, 773, 44
\bibitem[Warren et al.(2007)]{war07} Warren, S.~J., Hambly, N.~C., 
	Dye, S., et al.\ 2007, \mnras, 375, 213
\bibitem[Willott et al.(2005)]{wil05} Willott, C.~J., Delfosse, X.,
   Forveille, T., Delorme, P., \& Gwyn, S.~D.~J.\ 2005, \apj, 633, 630
\bibitem[Willott et al.(2007)]{wil07} Willott, C.~J., Delorme, P., 
	Omont, A., et al.\ 2007, \aj, 134, 2435
\bibitem[Willott et al.(2009)]{wil09} Willott, C.~J., Delorme, P., 
	Reyl{\'e}, C., et al.\ 2009, \aj, 137, 3541 
\bibitem[Willott et al.(2010)]{wil10} Willott, C.~J., Delorme, P., 
	Reyl{\'e}, C., et al.\ 2010, \aj, 139, 906
\bibitem[Wu et al.(2012)]{wu12} Wu, X.-B., Hao, G., Jia, Z., Zhang, Y., 
	\& Peng, N.\ 2012, \aj, 144, 49 
\bibitem[York et al.(2000)]{york00} York, D.~G., Adelman, J., 
	Anderson, J.~E., Jr., et al.\ 2000, \aj, 120, 1579
\end{thebibliography}
\end{document}